\documentclass[useAMS,usenatbib]{mn2e}
\input {psfig.sty}
\newcommand{\etal}{{et al.~}}

\newcommand{\Lsun}{\>{\rm L_{\odot}}}

\newcommand{\Msunh}{\>h^{-1}\rm M_\odot}

\newcommand{\beq}{\begin{equation}}
\newcommand{\eeq}{\end{equation}}

\newcommand{\rmd}{{\rm d}}

\newcommand{\lcen}{L_{\rm c}}
\newcommand{\avg}[1]{\langle #1 \rangle}
\newcommand{\avglogm}{\avg{\log M}(\lcen)}
\newcommand{\avgloglc}{\avg{\log \lcen}(M)}

\newcommand{\siglogm}{\sigma_{\log M}(\lcen)}

\newcommand{\drm}{{\rm d}}
\newcommand{\pdv}{{P (\Delta V)}}
\newcommand{\dv}{{\Delta V}}
\newcommand{\sigcen}{\sigma_{\log L}}
\newcommand{\sigsw}{\sigma_{\rm sw}}
\newcommand{\sighw}{\sigma_{\rm hw}}
\newcommand{\sigsat}{\sigma_{\rm sat}}
\newcommand{\avgsigsatsq}{\avg{\sigsat^2}}
\newcommand{\avnsat}{\avg{N_{\rm sat}}}
\newcommand{\avnsatm}{{\avnsat}_M}

\newcommand{\philsm}{\Phi_{\rm s}(L|M)}
\newcommand{\lc}{L_{\rm c}}

\newcommand{\pmlc}{P(M|\lc)}

\def\gtsima{$\; \buildrel > \over \sim \;$}
\def\ltsima{$\; \buildrel < \over \sim \;$}
\def\prosima{$\; \buildrel \propto \over \sim \;$}
\def\gsim{\lower.7ex\hbox{\gtsima}}
\def\lsim{\lower.7ex\hbox{\ltsima}}
\def\simgt{\lower.7ex\hbox{\gtsima}}
\def\simlt{\lower.7ex\hbox{\ltsima}}
\def\simpr{\lower.7ex\hbox{\prosima}}
\def\la{\lsim}

\def\lta{\la}

\newdimen\hssize
\newdimen\hdsize
\begin{document}
\setlength{\hbadness}{10000}

\title[Satellite Kinematics I]
       {Satellite Kinematics I: A New Method to Constrain
       the Halo Mass-Luminosity Relation of Central Galaxies}
\author[More, van den Bosch \& Cacciato]
       {Surhud More\thanks{International Max Planck Research
        School fellow \newline more@mpia.de},
        Frank C. van den Bosch,
        Marcello Cacciato\\
        Max Planck Institute for Astronomy, K\"onigstuhl 17, D-69117, 
        Heidelberg, Germany.}


\date{}

\maketitle

\label{firstpage}


\begin{abstract}
  Satellite kinematics can be used  to probe the masses of dark matter
  haloes of central galaxies. In  order to measure the kinematics with
  sufficient signal-to-noise, one uses the satellite galaxies of  a
  large  number  of central galaxies  stacked according to  similar
  properties  (e.g.,
  luminosity).   However, in general the relation
  between the luminosity of a central  galaxy and the mass of its host
  halo  will  have  non-zero  scatter.   Consequently,  this  stacking
  results in combining the  kinematics of satellite galaxies in haloes
  of  different masses,  which complicates  the interpretation  of the
  data.  In  this paper  we present an  analytical framework  to model
  satellite kinematics, properly accounting  for this scatter and
  for  various selection  effects.  We  show that  in the  presence of
  scatter in the halo mass-luminosity  relation, the commonly used
  velocity dispersion of satellite  galaxies can not be used  to infer
  a unique halo mass-luminosity relation.  In  particular, we
  demonstrate that there is  a   degeneracy  between  the   mean and
  the  scatter   of  the halo
  mass-luminosity relation.  We present a new technique that can break
  this   degeneracy,  and  which   involves  measuring   the  velocity
  dispersions  using two  different weighting  schemes: host-weighting
  (each central  galaxy gets the same weight)  and satellite-weighting
  (each central  galaxy gets  a weight proportional  to its  number of
  satellites).  The ratio  between the  velocity  dispersions obtained
  using  these two  weighting  schemes  is a  strong  function of  the
  scatter in  the halo mass-luminosity  relation, and can  thus be 
  used to infer a unique  relation between light and mass  from the
  kinematics of satellite galaxies.
\end{abstract}




\section{Introduction}
\label{sec:intro}

According to the current paradigm, the mass of  a  dark matter  halo
is believed to strongly  influence the process of galaxy  formation.
Hence, a  reliable determination  of the masses  of
dark matter haloes can provide important constraints on the physics of
galaxy formation.  Numerous methods are available  to probe the masses
of     dark   matter     haloes,       including    rotation    curves
\citep[e.g.,][]{Rubin1982},  gravitational  lensing,    either  strong
\citep[e.g.,][]{Gavazzi2007} or   weak \citep[e.g.][]{Mandelbaum2006},
X-ray        emission from     the    hot    intra-cluster      medium
\citep[e.g.,][]{Rykoff2008}, and a combined analysis of clustering and
abundances  of  galaxies \citep[e.g.,][]{Yang2003}.  In this  paper we
examine  another    powerful method, which   relies   on measuring the
kinematics  of satellite galaxies  in order  to infer  the mass of the
dark matter halo in which they orbit.

In general, a  dark matter halo  hosts one central galaxy, located  at
rest at the center of  the halo's potential  well, and a population of
satellite  galaxies orbiting  the halo.  The  line-of-sight (hereafter
los) velocity   dispersion  of these  satellite galaxies  reflects the
depth  of the halo's potential   well, and is   thus a measure for the
halo's  mass.   In the case  of rich  galaxy   clusters, the number of
satellite   galaxies can  be  sufficient  to  properly  sample the los
velocity    distribution     of       its    dark    matter       halo
\citep{Carlberg1996,Carlberg1997}.   In  less massive haloes, however,
the  number of (detectable)  satellite galaxies is generally too small
to obtain a reliable measure of the los velocity dispersion.  However,
under the assumption   that central galaxies with  similar  properties
(i.e., luminosity)  are hosted  by haloes  of  similar masses, one can
stack central galaxies and combine the kinematic information of
 their associated  satellites to improve the
statistics.   Pioneering   efforts  in this  direction   were  made by
\citet{Erickson1987},  \citet{Zaritsky1993}, \citet{Zaritsky1994}  and
\citet{Zaritsky1997}.  Although  these studies  were typically limited
to samples of less than 100  satellites, they nevertheless sufficed to
demonstrate  the existence of  extended    dark matter haloes   around
(spiral) galaxies.

More recently,  with the advent  of large homogeneous  galaxy redshift
surveys such as the Sloan Digital Sky Survey \citep[SDSS;][]{York2000}
and    the     Two    degree    Field     Galaxy    Redshift    Survey
\citep[2dFGRS;][]{Colless2001}, it  has become possible  to apply this
method    to    much   larger    samples    of   satellite    galaxies
\citep{McKay2002,Prada2003,Brainerd2003,van                         den
  Bosch2004,Conroy2007,Becker2007}.  These  studies all found  the los
velocity  dispersion  of satellite  galaxies,  $\sigma_{\rm sat}$,  to
increase  with the luminosity  of the  host (central)  galaxy, $L_{\rm
  c}$.  This  is in agreement  with the expectation that  more massive
haloes   host   more   luminous   centrals.   In   a   recent   study,
\citet{Norberg2008}  have shown  that  the quantitative  discrepancies
between  these  previous studies  mainly  owe  to  differences in  the
criteria  used to  select central  hosts and  their  satellites.  This
underscores the necessity for a careful treatment of selection effects
in order to extract reliable mass estimates from satellite kinematics.

Except for \citet{van den  Bosch2004}, all previous studies have  been
extremely     conservative  in    their   selection  of     hosts  and
satellites.  Consequently, despite the fact  that the redshift surveys
used  contain well in  excess of  100,000  galaxies, the final samples
only contained about $2000 - 3000$  satellite galaxies.  This severely
limits   the statistical    accuracy    of the   velocity   dispersion
measurements as well as the dynamic range in luminosity of the central
galaxies for  which halo masses  can be inferred.  The main motivation
for  using strict selection    criteria is to select only   `isolated'
systems,   with satellites that  can   be treated as tracer  particles
(i.e., their mass   does not cause  significant perturbations   in the
gravitational potential of their host galaxy).  Let $\pmlc$ denote the
conditional  probability distribution  that    a  central galaxy    of
luminosity $L_{\rm c}$  resides in a halo of  mass $M$. If the scatter
in $\pmlc$ is  sufficiently small, preferentially selecting `isolated'
systems should yield an unbiased estimate of $\langle M \rangle(L_{\rm
c})$,  which is the first  moment of $P(M|L_{\rm  c})$.  However, very
little is known about the actual amount of scatter in $P(M|L_{\rm c})$
and  different  semi-analytical  models  for  galaxy  formation   make
significantly different  predictions (see discussion in  Norberg \etal
2008).  If  appreciable, the   scatter will severely    complicate the
interpretation  of   satellite  kinematics,   and  may  even  cause  a
systematic    bias     (van  den  Bosch    \etal    2004;   More \etal
2008). Furthermore,   even if  the  scatter  is   small,  in practice,
satellites of central  galaxies  stacked in finite bins  of luminosity
are used to measure the kinematics.  If the satellite sample is small,
one has to resort to relatively large bins in order to have sufficient
signal-to-noise.   Therefore,  even  if the    distribution $\pmlc$ is
relatively narrow, this still implies mixing  the kinematics of haloes
spanning a relatively large range in halo masses.

In this paper we demonstrate  that whenever the scatter in $P(M|L_{\rm
  c})$ is  non-negligible, the $\sigma_{\rm  sat}(L_{\rm c})$ inferred
from the data has to be interpreted with great care. In particular, we
demonstrate that  there is a  degeneracy between the first  and second
moments of $P(M|L_{\rm c})$,  in that two distributions with different
$\langle M \rangle(L_{\rm c})$ and  different scatter can give rise to
the same $\sigma_{\rm sat}(L_{\rm c})$. Therefore, a unique $\langle M
\rangle(L_{\rm  c})$  cannot  be  inferred from  satellite  kinematics
without a  prior knowledge of the  second moment of  $P(M|L_{\rm c})$. 
However, not all  hope is lost. In fact, we  demonstrate that by using
two different  methods to  measure $\sigma_{\rm sat}(L_{\rm  c})$, one
can actually  break this degeneracy  and thus constrain both  the mean
and the scatter  of $P(M|L_{\rm c})$.  In this  paper we introduce the
methodology,  and   present  the  analytical   framework  required  to
interpret the data,  taking account of the selection  criteria used to
identify the  central host galaxies  and their satellites. In  More et
al. (2008;  hereafter Paper~II)  we apply this  method to the  SDSS to
infer both the mean and the scatter of $P(M|L_{\rm c})$, which we show
to be in good agreement  with the results obtained from clustering and
galaxy-galaxy   lensing  analyses.   In   addition,  in   Paper~II  we
demonstrate  that (i)  the  scatter  in $P(M|L_{\rm  c})$  can not  be
neglected,  especially not  at the  bright  end, and  (ii) the  strict
isolation criteria  generally used  to select centrals  and satellites
result  in  a  systematic  underestimate  of  the  actual  $\langle  M
\rangle(L_{\rm c})$.

This paper is organized  as follows.  In Section~\ref{sec:measure}, we
present two different schemes  to measure the velocity dispersion, the
satellite-weighting   scheme  and   the  host-weighting   scheme.   In
Section~\ref{sec:motiv},  we present  a toy  model which  serves  as a
basis  for   understanding  the  dependence   of  velocity  dispersion
estimates   on    the   different   parameters    of   interest.    In
Sections~\ref{sec:analest} and~\ref{sec:hod}  we refine our  toy model
by  including  selection  effects   and  by  using  a  realistic  halo
occupation distribution (HOD) model  for the central galaxies.  We use
these more  realistic models to  investigate how changes in  the halo
occupation  statistics   of  central  galaxies   affect  the  velocity
dispersion  of   satellite  galaxies,  and  we   demonstrate  how  the
combination of  satellite-weighting and host-weighting can  be used to
infer both the  mean and the scatter of  the mass-luminosity relation. 
We summarize our findings in Section~\ref{sec:summary}. Throughout
this paper $M$ denotes the halo mass in units of $\Msunh$.

\section{Weighting Schemes}
\label{sec:measure}

In order  to estimate dynamical  halo masses from satellite kinematics
one  generally proceeds  as  follows.  Using   a sample of   satellite
galaxies, one determines the distribution $\pdv$,  where $\Delta V$ is
the difference in the line-of-sight velocity of a satellite galaxy and
its  corresponding central host galaxy.   The  scatter in the
distribution $\pdv$ (hereafter   the velocity dispersion),  is then  
considered  to be an estimator of the depth  of the potential  well in
which the satellites orbit, and hence  of the halo  mass associated
with  the  central.  In order to  measure  the velocity dispersion 
as a function  of central galaxy  luminosity, $\sigma_{\rm 
sat}(L_{\rm c})$,  with sufficient signal-to-noise,  one has to
combine the  los velocity information of satellites  which belong to 
centrals of  the same luminosity, $\lc$. This procedure is influenced
by  two effects, namely {\it mass-mixing} and {\it
satellite-weighting}, which we now discuss in turn.

Mass-mixing refers to   combining the kinematics of satellites  within
haloes of  different masses.  The mass-luminosity  relation (hereafter
MLR) of central  galaxies can have  an appreciable scatter, i.e.,  the
conditional probability distribution  $\pmlc$ is not guaranteed to  be
narrow.  In this case, the   satellites used to measure $\sigma_{\rm
  sat}(L_{\rm c})$ reside in halo masses drawn from this distribution,
and $\sigma_{\rm  sat}(L_{\rm c})$  has to   be interpreted as an
average over $\pmlc$.

In most studies to date,  the technique used to measure $\sigma_{\rm
  sat}(L_{\rm c})$ implies satellite weighting. This can be elucidated
as  follows.   Let us   assume that one   stacks  $N_{\rm  c}$ central
galaxies, and that the $j^{\rm th}$ central  has $N_j$ satellites. The
total    number   of    satellites  $N_{\rm   sat}$    is    given  by
$\sum_{j=1}^{N_{\rm c}}   N_j$.   Let $\Delta  V_{ij}$ denote  the los
velocity difference between the $i^{\rm th}$ satellite and its central
galaxy $j$.  The  average velocity dispersion  of  the stacked system,
$\sigsw$, is such that
\begin{equation}
\label{sigswprev}
\sigsw^2 = \frac{\sum_{j=1}^{N_{\rm c}} \sum_{i=1}^{N_j} 
(\Delta V_{ij})^2}{\sum_{j=1}^{N_{\rm c}} N_j}  = 
{1 \over N_{\rm sat}} \sum_{j=1}^{N_{\rm c}} N_j \sigma^2_j \, .
\end{equation}
Here $\sigma_j$   is  the velocity dispersion in    the halo of the
$j^{\rm th}$ central galaxy. The velocity  dispersion measured in this
way is clearly a satellite-weighted average of the velocity dispersion
$\sigma_j$ around  each central galaxy \footnote{Note that the
velocity dispersion is always averaged in quadrature.}. Although not
necessarily directly using Eq.~(\ref{sigswprev}), most previous
studies have adopted this satellite-weighting scheme (McKay \etal
2002; Brainerd \& Specian 2003; Prada \etal 2003; Norberg \etal 2008).


In  principle, the satellite-weighting  can be undone by introducing a
weight $w_{ij} =  1/N_j$ for each  satellite-central  pair in the  los
velocity  distribution   \citep{van   den  Bosch2004,Conroy2007}.  The
resulting {\it host-weighted} average velocity dispersion, $\sighw$,
is such that
\begin{equation}
\sighw^2 = \frac{\sum_{j=1}^{N_{\rm c}} \sum_{i=1}^{N_j} w_{ij}
(\Delta V_{ij})^2}{\sum_{j=1}^{N_{\rm c}} w_{ij} N_j} =
{1 \over N_{\rm c}} \sum_{j=1}^{N_{\rm c}} \sigma^2_j \,,
\end{equation}
and it gives each halo an equal weight.

Consider a sample of  central and satellite galaxies with luminosities
$L   >    L_{\rm   min}$.     The   velocity   dispersions    in   the
satellite-weighting  and host-weighting  schemes  can be  analytically
expressed as follows:
\begin{equation}\label{sweqn}
\sigsw^2(\lcen) =
\frac{\int_{0}^{\infty} \pmlc \, \avnsatm \, \avgsigsatsq_M \, \rmd M }
     {\int_{0}^{\infty} \pmlc \, \avnsatm \, \rmd M }\,,
\end{equation}
\begin{equation}\label{hweqn}
\sighw^2(\lcen) =
\frac{\int_{0}^{\infty} \pmlc \, \avgsigsatsq_M \, \rmd M }
     {\int_{0}^{\infty} \pmlc \, \rmd M }\,.
\end{equation}
Here $\avnsatm$  denotes the  average number of  satellites with  $L >
L_{\rm  min}$ in  a  halo of  mass  $M$, and  $\avgsigsatsq_M$ is 
the square of the los velocity dispersion of satellites averaged over
the entire halo.

Consider a MLR of central galaxies  that has no scatter, i.e. $\pmlc =
\delta(M-M_0)$,  where  $M_0$ is  the  halo  mass  for a  galaxy  with
luminosity $\lc$.  In this case  both schemes give an equal measure of
the    velocity   dispersion,   i.e.,    $\sigsw^2   =    \sighw^2   =
\avgsigsatsq_{M_0}$. Most studies to  date have assumed the scatter in
$\pmlc$  to  be  negligible,  and  simply  inferred  an  average  MLR,
$M_0(\lc)$     using      $\sigsw^2(\lc)     =     \avgsigsatsq_{M_0}$
(McKay \etal 2002; Brainerd \& Specian 2003; Prada \etal 2003;
Norberg \etal 2008). However, as shown  in
\citet{van den Bosch2004}, and as  evident from  the above equations 
(\ref{sweqn})  and (\ref{hweqn}), whenever  the  scatter in $\pmlc$ is
non-negligible,  $\sigsw^2(\lc)$  and  $\sighw^2(\lc)$  can     differ
significantly\footnote{Note  that $\sigsw^2     \ne   \sighw^2$ is   a
sufficient but not a  necessary condition to  indicate the presence of
scatter in $\pmlc$; after  all, if $\avnsatm$ does  not depend on mass
then $\sigsw^2=\sighw^2$ independent  of the amount of scatter.} (see
also Paper II).

In this paper, we show that ignoring the scatter in the MLR of central
galaxies can  result in appreciable errors   in the inferred mean
relation  between mass and  luminosity.   We show, though,  that these
problems can  be  avoided by simultaneously  modeling $\sigsw^2(\lc)$
and $\sighw^2(\lc)$. In  particular, we demonstrate  that the ratio of
these  two quantities can be used  to determine  the actual scatter in
the MLR of central galaxies.

\section{Toy model}
\label{sec:motiv}

In the previous  section, we have shown that  both $\sigsw^2(\lc)$ and
$\sighw^2(\lc)$  can  be  analytically   expressed  in  terms  of  the
probability  function, $\pmlc$,  the satellite  occupation, $\avnsatm$,
and the kinematics of the satellite galaxies within a halo of mass M
specified by $\avgsigsatsq_M$. In   fact,   the    inversion   of  
equations   ($\ref{sweqn}$)   and ($\ref{hweqn}$) presents an
opportunity to constrain $\pmlc$ using the observable $\sigsw^2$ and
$\sighw^2$.  In this section we use a simple toy  model  to
demonstrate that  the  combination  of $\sigsw^2$  and
$\sighw^2$ can be  used to constrain the first  two moments (i.e., the
mean and the scatter) of $\pmlc$.

For  convenience, let     us   assume that  $\pmlc$  is   a  lognormal
distribution 
\begin{equation}\label{pmlc}
P(M|\lcen)\,\rmd M = {1\over\sqrt{2\pi\sigma^2_{\ln M}}}\exp\left[-\left(
{\ln (M/M_0) \over \sqrt{2 \, \sigma^2_{\ln M}}}\right)^2\right]
{\rmd M \over M}\,.
\end{equation}
Here $M_0$ is a characteristic mass scale which obeys
\begin{equation}\label{lnmlcreln}
\ln M_0=\int_{0}^{\infty} \pmlc \ln M \rmd M
        = \avg{\ln M}\,,
\end{equation}
and  $\sigma^2_{\ln M}$ reflects the scatter  in halo  mass at a fixed
central luminosity and is given by
\begin{equation}\label{siglnmlcreln}
\sigma^2_{\ln M} = \int_{0}^{\infty} \pmlc (\ln M - \ln M_0 )^2
\rmd M .
\end{equation}

In addition,  let us assume that  both $\avgsigsatsq_M$ and $\avnsatm$
are simple power laws,
\begin{equation}\label{avns}
\avnsatm = \tilde{N} \left({M \over 10^{12} }\right)
^{\alpha}\,,
\end{equation}
\begin{equation}\label{avss}
{\avgsigsatsq}_M = \tilde{S}^2 \left({M \over 10^{12} }\right)
^{\beta}\,.
\end{equation}
with  $\alpha$  and $\beta$  two constants,   $\tilde{N}$  the average
number  of satellites  in    a halo of  mass  $10^{12}   \Msunh$,  and
$\tilde{S}$ the corresponding los velocity dispersion.

Substituting     Eqs.~(\ref{pmlc}), (\ref{avns})  and  (\ref{avss}) in
Eqs.~(\ref{sweqn}) and (\ref{hweqn}) yields
\begin{equation}\label{sigswtoy}
 \sigsw^2(\lcen) = \tilde{S}^2 \left({M_0 \over
10^{12} }\right)^\beta \exp\left[\frac{\sigma_{\ln
M}^2\beta^2}{2} 
 \left(1 + 2{\alpha\over\beta}\right)\right]\,,
\end{equation}
\begin{equation}\label{sighwtoy}
 \sighw^2(\lcen) = \tilde{S}^2 \left({M_0 \over
10^{12} }\right)^\beta \exp\left[\frac{\sigma_{\ln
M}^2\beta^2}{2} 
 \right]\,.
\end{equation}
The velocity   dispersions $\sigsw(\lcen)$  and    $\sighw(\lcen)$
depend on both $M_0$ and  $\sigma_{\ln M}$, elucidating the degeneracy
between   the  mean mass  $M_0(\lcen)$   and the  scatter $\sigma_{\ln
  M}(\lcen)$  of the  distribution  $\pmlc$.  In particular, if   only
$\sigsw(\lcen)$ or $\sighw(\lcen)$  is measured, one cannot deduce
$M_0(\lcen)$ without having an  independent  knowledge of the  scatter
$\sigma_{\ln M}(\lcen)$. However, the latter  can be inferred from the
{\it ratio}  of  the satellite-weighted to  the host-weighted velocity
dispersion. In particular, in the case of our toy model,
\begin{equation}\label{scatter}
\sigma_{\ln M}^2 = {1\over\alpha\beta} \ln \left(
\frac{\sigsw^2}{\sighw^2} \right)
\end{equation}
Thus, by measuring  {\it both} $\sigsw(\lcen)$ and $\sighw(\lcen)$
one  can  determine both  $M_0(\lcen)$  and  its scatter  $\sigma_{\ln
  M}(\lcen)$,  provided that  the constants  $\alpha$ and  $\beta$ are
known.  Since virialized dark matter  haloes all have the same average
density    within    their     virial    radii,    $\beta    =    2/3$
\citep[e.g.][]{Klypin1999,van  den Bosch2004}.  Previous  studies have
obtained constraints on $\alpha$ that cover the range $0.7 \lta \alpha
\lta           1.1$           \citep[e.g.][]{Yang2005,van          den
  Bosch2007,Tinker2007,Yang2007}.  Since $\sigma_{\ln M}^2 \propto
\alpha^{-1}$, this uncertainty directly translates into an uncertainty
of the inferred  scatter. Therefore, in Paper II,  we  do not use  the
constraints on $\alpha$ available in the literature  to infer the mean
and scatter of the MLR from real data.   Instead, we treat $\alpha$ as
a free parameter and use the average number  of observed satellites as
a function of the luminosity of central as an additional constraint.

\section{Selection effects}
\label{sec:analest}

The toy  model presented  in   the previous section  illustrates  that
measurements of the satellite-weighted and host-weighted kinematics of
satellite galaxies can be used  to infer the mean  and scatter of  the
MLR of central galaxies, $\pmlc$. However, in practice one first needs
a method to   select central galaxies   and satellites from a   galaxy
redshift survey. In general, central  galaxies are selected to be  the
brightest  galaxy  in some  cylindrical volume  in redshift space, and
satellite galaxies are defined as those galaxies that are fainter than
the  central  by a certain   amount  and located within  a cylindrical
volume  centered on the  central. In  this  section we show  how these
selection  criteria impact on $\sigsw^2$  and $\sighw^2$, and how this
can be accounted for in the analysis.

No selection criterion  is perfect, and some galaxies will
be  selected  as  centrals,  while  in  reality  they  are  satellites
(hereafter  `false centrals').   In  addition, some  galaxies will  be
selected as satellites of a  certain central, while in reality they do
not reside in the same  halo as the central (hereafter `interlopers'). 
The  selection criteria  have to  be tuned  in order  to  minimize the
impact  of these  false centrals  and interlopers.   Here we  make the
assumption that interlopers can be  corrected for, and that the impact
of false centrals is  negligible.  Using mock galaxy redshift surveys,
van den  Bosch \etal (2004) have  shown that one  can devise adaptive,
iterative selection criteria that  justify these assumptions (see also
Paper II).  Here  we focus on the impact  of these iterative selection
criteria on the satellite kinematics in the absence of interlopers and
false centrals. Our analytical treatment for selection effects
follows the one presented in \citet{van den Bosch2004} except for the
inclusion of an extra selection effect. For completeness, we outline
this treatment below.

In  general,   satellite  galaxies  are  selected  to   lie  within  a
cylindrical volume  centered on its  central galaxy, and  specified by
$R_{\rm p} <  R_{\rm s}$ and $|\dv|  < (\dv)_{\rm s}$.  Here  $R_{\rm
p}$ is
the physical separation  from the central galaxy projected  on the sky
and  $\dv$  is  the  los  velocity difference  between a satellite 
and its central.  Usually, $(\dv)_{\rm s}$ is chosen sufficiently
large, so that  it  does  not  exclude  true satellites  from  being 
selected. However, in the adaptive, iterative selection criteria of
\citet{van den Bosch2004},  which we  will use in  our companion
paper  II, the
aperture  radius is  tuned so  that $R_{\rm  s} \simeq  0.375\, r_{\rm
  vir}$, where $r_{\rm  vir}$ is the virial radius  of the dark matter
halo hosting  the central-satellite pair.  This  means that $\avnsatm$
and $\avgsigsatsq_M$  in Eqs.~(\ref{sweqn}) and  (\ref{hweqn}) need to
be replaced by $\avnsat_{{\rm ap},M}$ and $\avgsigsatsq_{{\rm ap},M}$,
respectively.  Here  $\avnsat_{{\rm ap},M}$  is the average  number of
satellites in  a halo of  mass $M$ that  lie within the  aperture, and
$\avgsigsatsq_{{\rm   ap},M}$  is   the square of the los velocity
dispersion of satellite galaxies averaged over the aperture.

The  number of  satellites present  within  the aperture,
$\avnsat_{{\rm ap},M}$,  is related to the number  of satellites given
by the halo occupation statistics, $\avnsatm$, via
\begin{equation}\label{nsatproj}
  \avnsat_{{\rm ap},M} = \left\{ 
  \begin{array}{ll}
    f_{\rm cut} \, \avnsatm & \mbox{if $R_{\rm s} < r_{\rm vir}$} \\
    \avnsatm                & \mbox{if $R_{\rm s} \geq r_{\rm vir}$}
  \end{array}\right.
\end{equation}
with
\begin{equation}\label{fcut}
f_{\rm cut} = {4 \pi \over \avnsatm} \int_0^{R_{\rm s}} R \, \rmd R 
\int_{R}^{r_{\rm vir}} n_{\rm sat}(r|M) \, 
{r \, \rmd r \over \sqrt{r^2 - R^2}} \,.
\end{equation}
Here   $n_{\rm sat}(r|M)$  is   the  number  density distribution   of
satellites within a halo of mass $M$, which is normalized so that
\begin{equation}
\avnsatm = 4 \pi \int_0^{r_{\rm vir}} n_{\rm sat}(r|M) \, r^2 \, \rmd r\,.
\end{equation}

Under  the assumption that  the satellites  are in  virial equilibrium
within  the dark  matter halo,  and  that the  velocity dispersion  of
satellite galaxies within a given  halo is isotropic, the los velocity
dispersion  of satellites  within the  cylindrical aperture  of radius
$R_{\rm s}$ is given by
\begin{eqnarray}\label{sigproj}
\avgsigsatsq_{{\rm ap},M} & = & {4 \pi \over \avnsat_{{\rm ap},M}}
\int_0^{R_{\rm s}} \rmd R \, R \nonumber \\
& & \int_R^{r_{\rm vir}} n_{\rm sat}(r|M) \, \sigma_{\rm sat}^2(r|M) \, 
{r \, \rmd r \over \sqrt{r^2 - R^2}}\,.
\end{eqnarray}
Here  $\sigma_{\rm  sat}(r|M)$  is the  local,  one-dimensional
velocity dispersion which is related to the potential $\Psi$ of the
dark matter halo via the Jeans equation
\begin{equation}\label{sigr}
\sigsat^2(r|M) = {1\over n_{\rm sat}(r|M)} \int_r^{\infty}
n_{\rm sat}(r'|M) \, \frac{\partial \Psi}{ \partial r'}(r'|M) \, \rmd r'.
\end{equation}
The radial  derivative of the  potential $\Psi$ represents  the radial
force and is given by
\begin{equation}\label{dpsidr}
 \frac{\partial \Psi}{ \partial r}(r|M) = {4 \pi G \over r^2 } 
\int_0^{r} \rho(r'|M) \, r'^2 \, \rmd r' \,,
\end{equation}
with $\rho(r|M)$  the density  distribution of a  dark matter  halo of
mass $M$.  The assumptions  of virial equilibrium and orbital isotropy
are supported  by results from  numerical simulations which  show that
dark matter subhaloes  (and hence satellite galaxies) are  in a steady
state equilibrium  within the  halo and that  their orbits  are nearly
isotropic  at  least  in  the  central  regions  \citep{Diemand2004}.  
Furthermore,   \citet{van  den   Bosch2004}  have   demonstrated  that
anisotropy has a negligible  impact on the average velocity dispersion
within the selection aperture.

Finally, there is one other  effect of the selection criteria to be
accounted  for which has not been considered in \citet{
van den Bosch2004}. When selecting central-satellite pairs, only
those centrals  are selected  with at least  one satellite  inside the
search  aperture. This  has an  impact on  the  host-weighted velocity
dispersions that  needs to be  accounted for.  The probability  that a
halo of mass $M$,   which  on  average   hosts  $\avnsat_{{\rm ap},M}$
satellites  within  the  aperture  $R_s$,  has $N_{\rm  sat}  \geq  1$
satellites within the aperture, is given by
\begin{eqnarray}\label{poisson}
P(N_{\rm sat} \geq 1) &=& 1 - P(N_{\rm sat}=0) \nonumber \\
     & = & 1 - {\rm exp}\left[-\avnsat_{{\rm ap},M}\right] \nonumber \\
& \equiv & {\cal P}(\avnsat_{{\rm ap},M}).
\end{eqnarray}
Here, for the second equality, we have assumed Poisson statistics for
the satellite occupation numbers. Note that, in the
satellite-weighting scheme, haloes that have zero satellites, by
definition, get zero weight. Therefore only the host-weighted velocity
dispersions need to be corrected for this effect.

Thus,  in  light  of  the selection  effects,  Eqs.~(\ref{sweqn})  and
(\ref{hweqn}) become
\begin{equation}\label{sweqnvol}
\sigsw^2(\lc) =\frac{ \int_{0}^{\infty}
\pmlc \, \avnsat_{{\rm ap},M} \, \avgsigsatsq_{{\rm ap},M} \, \rmd M }
{ \int_{0}^\infty \pmlc \, \avnsat_{{\rm ap},M} \, \rmd M },
\end{equation}
and
\begin{equation}\label{hweqnvol}
\sighw^2(\lc) = \frac{ \int_{0}^{\infty} \pmlc \,
{\cal P}(\avnsat_{{\rm ap},M}) \, \avgsigsatsq_{{\rm ap},M} \, \rmd M} 
{ \int_{0}^\infty \pmlc \, {\cal P}(\avnsat_{{\rm ap},M}) \, \rmd M }.
\end{equation}
Note that ${\cal P}(\avnsat_{{\rm ap},M}) \simeq \avnsat_{{\rm ap},M}$
when $\avnsat_{{\rm ap},M} \rightarrow  0$. This implies that $|\sigsw
- \sighw| \rightarrow 0$ for  faint centrals (i.e.  when $\lc$ becomes
comparable to the luminosity limit of the survey).

\section{More Realistic Models}
\label{sec:hod}

Using the methodology described above, we now illustrate how satellite
kinematics can  be used to constrain  the mean and the  scatter of the
MLR  of central  galaxies,  $\pmlc$.  We improve  upon  the toy  model
described in Section~\ref{sec:motiv}  by considering a realistic model
for the  halo occupation statistics  and take the impact  of selection
criteria into account.

As is evident from the discussion in the previous section, calculating
$\sigsw^2(\lc)$ and $\sighw^2(\lc)$ requires the following input:
\begin{itemize}
 \item the density distributions of dark matter haloes, $\rho(r|M)$
 \item the number density distribution of satellites, $n_{\rm sat}(r|M)$
 \item the halo occupation statistics of centrals, $\pmlc$.
\end{itemize}

We assume that  dark matter haloes  follow the NFW \citep{Navarro1997}
density distribution
\begin{equation}\label{rhor}
\rho(r|M) = {M \over 4\pi r_s^3 \mu(c)} 
\left( \frac{r}{r_{\rm s}}\right)^{-1} \left( 1 + \frac{r}{r_{\rm s}}
\right)^{-2}\,.
\end{equation}
Here,  $r_{\rm s}$  is  a  characteristic scale  radius,  $c =  r_{\rm
  vir}/r_{\rm s}$ is the halo's concentration parameter, and
\begin{equation}
\mu(x) \equiv \ln (1+x) - \frac{x}{1+x}\,.
\end{equation}
Throughout  we  use the  relation   between   $c$ and   $M$  given  by
\citet{Maccio2007}.
\begin{figure}
\centerline{\psfig{figure=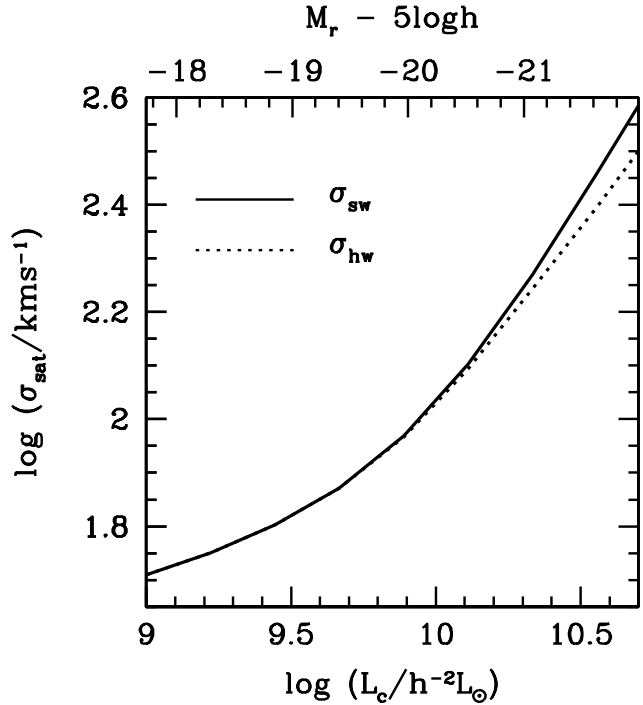,width=\hssize}}
\caption{The satellite-weighted ($\sigsw$) and host-weighted
  ($\sighw$) velocity dispersions of  satellite galaxies for model G1. 
  Note that $\sigsw(\lc) >  \sighw(\lc)$ at the bright end, indicating
  that the  MLR of central galaxies, $P(M|\lc)$,  has a non-negligible
  amount of scatter.}
\label{fig:hwswfid}
\end{figure}

We assume that satellite  galaxies are spatially unbiased with respect
to  the   dark  matter  particles,   so  that  their   number  density
distribution  is  given  by  Eq.~(\ref{rhor})  with  $M$  replaced  by
$\avnsatm$. Note that this is a fairly simplistic assumption. We
address the issue of potential spatial antibias of satellite galaxies
in Paper II.

Substituting $\rho(r|M)$ and $n_{\rm sat}(r|M)$ in Eqs.~(\ref{dpsidr})
and (\ref{sigr}) gives
\begin{equation}\label{sig1dm}
\sigma^2_{\rm sat}(r|M) =  {c \, V^2_{\rm vir} \over \mu(c)} \,
\left({r \over r_{\rm s}}\right) \, \left(1 + {r \over r_{\rm s}}\right)^{2} \,
\int_{r/r_{\rm s}}^{\infty} {\mu(x) \, \rmd x \over x^3 \, (1+x)^2}\,,
\end{equation}
where  $V_{\rm vir}   =  (G M /r_{\rm   vir})^{1/2}$   is the circular
velocity at $r_{\rm vir}$.

The  final ingredient  is a  realistic model  for the  halo occupation
statistics of  centrals and  satellites.  To that  extent, we  use the
conditional  luminosity  function (CLF)  presented  in Cacciato  \etal
(2008). The CLF, denoted by  $\Phi(L|M) \rmd L$, specifies the average
number of  galaxies with  luminosities in the  range $L \pm  \drm L/2$
that reside in  a halo of mass $M$, and is  explicitly written as the
sum of the  contributions due to central and  satellite galaxies, i.e. 
$\Phi(L|M) =  \Phi_{\rm c}(L|M) + \Phi_{\rm s}(L|M)$.   From this CLF,
the  probability  distribution  $\pmlc$  follows from  Bayes'  theorem
according to
\begin{equation}\label{bayes}
\pmlc = {\Phi_{\rm c}(\lc|M) \, n(M) \over
\int_0^{\infty} \Phi_{\rm c}(\lc|M) \, n(M) \, \rmd M}\,,
\end{equation}
with  $n(M)$ the  halo  mass  function, while  the  average number  of
satellites with $L \geq L_{\rm min}$ in a halo of mass $M$ is given by
\begin{equation}
\avnsatm = \int_{L_{\rm min}}^{\infty} \philsm \, \rmd L\,.
\end{equation}
\begin{table}
\caption{Different models for the HOD of centrals}
\begin{tabular}{lccccc}
\hline
Model & $\sigcen$ & $\gamma_1$ & $\gamma_2$ & $L_0$ & $M_1$ \\
\hline \hline
G1 & 0.14 & 3.27 & 0.25 & 9.94 & 11.07 \\
G2 & 0.25 & 3.27 & 0.25 & 9.94 & 11.07 \\
G3 & 0.14 & 1.80 & 0.40 & 9.80 & 11.46 \\
\hline
\end{tabular}
\medskip
\begin{minipage}{\hssize}
  Three different models describing the MLR of centrals used to
  predict $\sigsw(\lc)$ and $\sighw(\lc)$.
\end{minipage}
\label{table2}
\end{table}
\begin{figure*}
\centerline{\psfig{figure=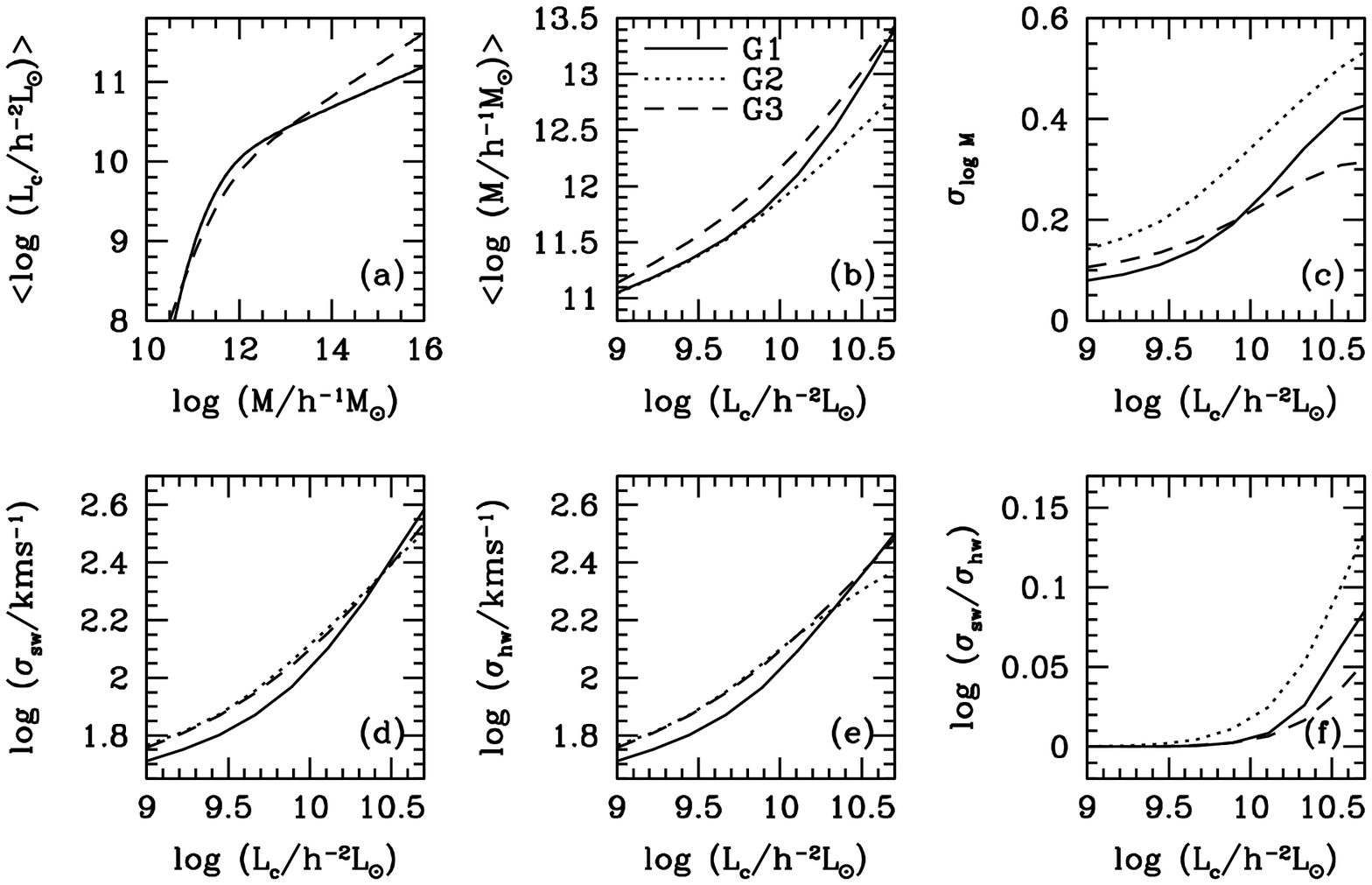,width=0.95\hdsize}}
\caption{Comparison of   three  models with   different  HODs for  the
  central galaxies. In all panels  the solid line corresponds to model
  G1, the dotted line to model G2 and the dashed line to model G3 (see
  Table~1  for  the  parameters).    Panels  (a),  (b)  and  (c)  show
  $\avgloglc$,  $\avglogm$ and  $\siglogm$, respectively.   Panels (d)
  and  (e)  show the  predicted  satellite-weighted and  host-weighted
  velocity dispersions as function  of luminosity, and panel (f) shows
  the logarithm of  the ratio between $\sigsw$ and  $\sighw$. See text
  for a detailed discussion.}
\label{fig:hwswpt}
\end{figure*}

The parametric forms  for $\Phi_{\rm c}(L|M)$ and $\Phi_{\rm  s}(L|M)$
are motivated by the  results of \citet[][hereafter  YMB08]{Yang2008},
who  determined the     CLF   from the    SDSS  group    catalogue  of
\citet{Yang2007}.  In  particular, $\Phi_{\rm  c}(L|M)$ is  assumed to
follow a log-normal distribution
\begin{equation}
\Phi_{\rm c}(L|M) \rmd L = \frac{\log{\rm e}}{\sqrt{2\pi} \, \sigcen} 
\exp\left( -\left[ {\log(L/L_{c}^*) \over {\sqrt{2} \sigcen}}
\right]^2 \right)\, {\rmd L \over L}\,,
\end{equation}
with $\sigcen$ a free parameter that we take to be independent of halo
mass, and
\begin{equation}\label{meanlcen}
\lc^*(M) = L_0 {(M/M_1)^{\gamma_1} \over \left[ 1 + (M/M_1)
\right]^{\gamma_1-\gamma_2}}
\end{equation}
which has four additional  free parameters: two slopes, $\gamma_1$ and
$\gamma_2$, a  characteristic halo  mass, $M_1$, and  a normalization,
$L_0$.  Note  that, $\lc^* \propto  M^{\gamma_1}$ for $M \ll  M_1$ and
$\lc^* \propto  M^{\gamma_2}$ for $M \gg M_1$.   Cacciato \etal (2008)
constrained  the free  parameters, $\sigcen$,  $\gamma_1$, $\gamma_2$,
$M_1$  and   $L_0$,  by  fitting  the  SDSS   luminosity  function  of
\citet{Blanton2003}  and the  galaxy-galaxy correlation  lengths  as a
function  of luminosity from \citet{Wang2007}.  The resulting best-fit
parameters are listed in the  first row of Table~1, and constitute our
fiducial  model  G1.  We  also  consider  two  alternative models  for
$\Phi_{\rm c}(L|M)$,  called G2  and G3, the  parameters of  which are
also listed in Table~1. For  $\Phi_{\rm s}(L|M)$ we adopt the model of
Cacciato  \etal (2008)  throughout,  without any  modifications: i.e.  
models G1, G2, and G3 only differ in $\pmlc$ and have the same $n_{\rm
  sat}(r|M)$.

Having  specified  all  necessary  ingredients,  we  now  compute  the
satellite  weighted  and host-weighted  satellite  kinematics for  our
fiducial  model G1 using  Eqs.~(\ref{sweqnvol}) and  (\ref{hweqnvol}). 
The   results    are   shown   as   solid   and    dotted   lines   in
Fig.~\ref{fig:hwswfid},  where  we have  adopted  a minimum  satellite
luminosity of  $L_{\rm min} =  10^9 h^{-2} \Lsun$.  At  the faint-end,
the velocity  dispersions $\sigsw$ and $\sighw$ are  equal, this
simply reflects the fact that $\avnsatm \rightarrow 0$ if  $\lc
\rightarrow L_{\rm min}$.  At the bright end, though, the non-zero 
scatter in $\pmlc$ causes the difference between $\sigsw$ and $\sighw$
to  increase systematically with increasing $\lc$.  This  is a generic
trend for any realistic halo occupation  model (see also van den Bosch
et al. 2004).

\begin{figure}
\centerline{\psfig{figure=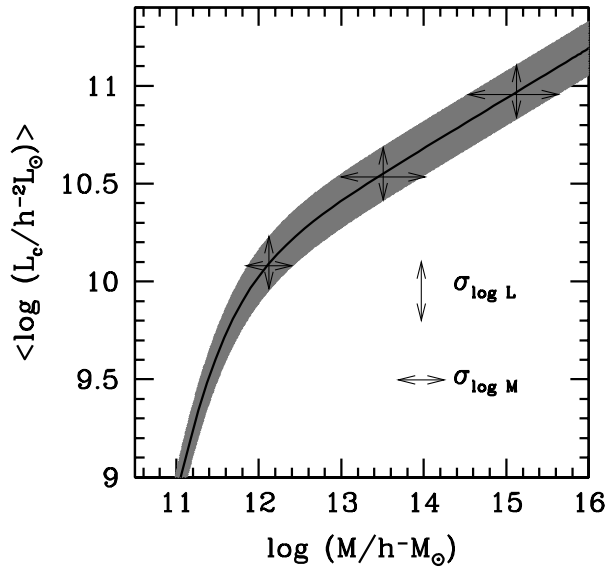,width=0.95\hssize}}
\caption{Illustration of the MLR of central galaxies. The solid black 
  line indicates  the mean of  the $\lc$-$M$ relation, while  the gray
  scale  region reflects  the scatter. In  this  particular case the
  scatter in $P(\lc|M)$ (indicated by vertical arrows) is taken  to be
  constant  with halo mass. Note,
  though,  that  the  scatter   in  $P(M|\lc)$  (indicated  by
  horizontal arrows)  increases with increasing $\lc$;  this simply
  is due to
  the  fact that the slope  of the mean  $\lc$-$M$ relation becomes
  shallower with increasing halo mass.}
\label{fig:scatter}
\end{figure}

The upper panels of Fig.~\ref{fig:hwswpt} show the mean and scatter of
the MLR  of central  galaxies in models  G1 (solid lines),  G2 (dotted
lines) and G3 (dashed lines). Panel (a) plots $\avgloglc = \log(L_{\rm
  c}^{*})$,   which   reveals  the   double   power-law  behavior   of
Eq.~(\ref{meanlcen}), panel (b) shows the inverse relation,
%
\begin{equation}\label{averlogm}
\avglogm = \int_{0}^{\infty} \pmlc \log M \, \rmd M\,.
\end{equation}
and panel (c) shows the scatter in the MLR, $\siglogm$, deduced by
using
\begin{equation}
\sigma^2_{\log M} = \int_{0}^{\infty} \pmlc \left[\log M -
\avglogm\right]^2
\, \rmd M\,.
\end{equation}
Note that $\siglogm$ increases  with increasing $\lc$, even though the
scatter $\sigcen$ is constant with halo mass.  This simply owes to
the fact that the slope of $\avgloglc$ becomes shallower with
increasing $\lc$, as illustrated in Fig.~\ref{fig:scatter}.

The  comparison  between models  G1 and G2  illustrates the  effect of
changing  the scatter $\sigcen$  in $\Phi_{\rm  c}(L|M)$.  Both models
have exactly the same $\avgloglc$  (the solid line overlaps the dotted
line in  panel a).   However, because  the scatter  $\sigcen$ in G2 is
larger than in G1 (see Table~1), the $\avglogm$ of G2 is significantly
lower than that of G1 at the bright end  ($\sim$ 0.5 dex at the bright
end).  This is due to the shape of the halo mass function.  Increasing
the    scatter adds  both  low   mass  and  high  mass  haloes  to the
distribution $\pmlc$ (cf.  Eqs.~[\ref{bayes}] and  ~[\ref{averlogm}]),
and the overall change  in the average halo mass  depends on the slope
of the halo mass function.  Brighter galaxies live  on average in more
massive haloes  where    the  halo  mass   function is  steeper.    In
particular, when  the halo mass range  sampled by $\pmlc$  lies in the
exponential tail of the halo mass function, an increase in the scatter
adds many more low mass haloes than massive haloes, causing a shift in
the average   halo mass  towards  lower  values.   On the other  hand,
fainter galaxies  live in less massive  haloes, where the slope of the
halo mass function is much  shallower.  Consequently, a change in  the
scatter does not   cause an appreciable  change in   the average mass.
Finally,  as  expected, the scatter in  the  MLR, $\siglogm$, in G2 is
higher than for G1 at all luminosities (see panel c).

Panels (d) and  (e)    of Fig.~\ref{fig:hwswpt} show   the  analytical
predictions  for $\sigsw(L_c)$ and  $\sighw(L_c)$, respectively.  Note
that  models  G1 and    G2  predict  satellite  kinematics  that   are
significantly different (which  can be distinguished given the typical
measurement errors in  Paper II),  even  though both have exactly  the
same  $\avgloglc$.  In particular, model   G2 predicts larger $\sigsw$
and $\sighw$ at the faint end, but  lower $\sigsw$ and $\sighw$ at the
bright end.   The  trend at  the faint is  due  to the  fact  that the
scatter $\siglogm$ is higher in G2 than in G1. Quantitatively, this is
evident from Eqs.~(\ref{sigswtoy})    and~(\ref{sighwtoy}),      which
demonstrate that   both the   satellite  weighted and    host weighted
satellite kinematics increase  with increasing scatter.  At the bright
end, however, the  drastic decrease in $\avglogm$  for G2 with respect
to G1 overwhelms this boost  and causes  $\sigsw$  and $\sighw$ to  be
lower in G2.

Now consider  model G3. This model  has the same amount  of scatter as
model     G1,     but     we     have     tuned     its     parameters
$(\gamma_1,\gamma_2,M_1,L_0)$ that describe  $\avgloglc$ such that its
$\sigsw(\lc)$ closely matches that of  model G2 (the dotted and dashed
curves in panel (d) are almost overlapping). As is evident from panels
(a)-(c), though,  the MLR  of G3 is  very different  from that of  G2. 
Note that  the higher values of  $\avglogm$ for G3  are compensated by
its  lower  values of  $\siglogm$,  such  that the  satellite-weighted
kinematics  are  virtually identical.   This  clearly illustrates  the
degeneracy  between the  mean  and the  scatter  of the  MLR: One  can
decrease the  mean of  the MLR  and yet achieve  the same  $\sigsw$ by
increasing the scatter  of the MLR. It also  shows that $\sigsw$ alone
does not yield sufficient information to uniquely constrain the MLR.

Note, though, that although $\sigsw$ is the same for models G2 and G3,
their host-weighted satellite kinematics, $\sighw(\lc)$, are different
at the bright end.  In fact, the  ratios $\sigsw/\sighw$ for models G2
and G3 are  clearly different.  The  logarithm of this ratio, shown in
panel (f), follows the  same trend as  $\siglogm$, i.e.  it is  higher
for model G2 than for  G3.  This is  in agreement with our toy  model,
according  to which  the  ratio  $\sigsw/\sighw$  increases with   the
scatter  $\siglogm$ (cf.  Eq.~[\ref{scatter}]).  This illustrates once
again that  the {\it combination} of $\sigsw$  and $\sighw$ allows one
to constrain both the mean and the scatter of the MLR simultaneously.

\section{Summary}
\label{sec:summary}

The kinematics of satellite galaxies is a powerful probe of the masses
of the dark matter   haloes  surrounding central galaxies.  With   the
advent of large, homogeneous redshift surveys,  it has become possible
to probe  the   mass-luminosity relation  (MLR)    of central galaxies
spanning a  significant  range in  luminosities.  Unfortunately, since
most centrals only host  a  few satellite galaxies with   luminosities
above the flux  limit of the redshift  survey, one generally needs  to
stack a large number of central galaxies within a given luminosity bin
and combine the velocity  information of their satellites. Because  of
the finite bin-width, and because  the MLR has intrinsic scatter, this
stacking results in combining the kinematics  of satellite galaxies in
haloes of different masses, which   complicates the interpretation  of
the  data. Unfortunately, most   previous  studies have  ignored  this
issue,  and made  the oversimplified assumption  that  the scatter  is
negligible.

Using realistic models for the halo  occupation statistics, and taking
account of selection  effects,    we have demonstrated  a   degeneracy
between the mean  and the scatter of the  MLR: one can change the mean
relation between halo mass, $M$, and central galaxy luminosity, $\lc$,
and simultaneously change the scatter around  that mean relation, such
that    the   observed   satellite kinematics,   $\langle  \sigma_{\rm
sat}\rangle(\lc)$, are unaffected.

We have also presented a new technique to break this degeneracy, based
on measuring the  satellite  kinematics using two  different weighting
schemes: host-weighting (each central galaxy gets the same weight) and
satellite weighting (each central galaxy gets a weight proportional to
its number of satellites).  In  general, for central galaxies close to
the magnitude limit  of the survey,  the average number  of satellites
per host is  close   to  zero, and the  satellite-weighted    velocity
dispersion,  $\sigsw$,  is  equal   to   the host-weighted    velocity
dispersion,   $\sighw$.  This is because   only those centrals with at
least one satellite are used  to measure the satellite kinematics. For
brighter centrals, however, $\sigsw > \sighw$  and the actual ratio of
these  two  values  is larger    for  MLRs with    more scatter   (see
Eq.~[\ref{scatter}] and panels   c  anf f   of Fig.~\ref{fig:hwswpt}).
Hence,  the combination of   $\sigsw(\lc)$ and $\sighw(\lc)$  contains
sufficient information  to constrain both the mean  and the scatter of
the MLR of central galaxies. In our companion paper (More \etal 2008),
we apply this method to the SDSS, and show that  the amount of scatter
inferred  from   the data is    in   excellent agreement with   other,
independent constraints.  In the companion paper,  we also address the
issues of  measurement errors, sampling  effects and interlopers.

In  a  recent study,  \citet{Becker2007}  analyzed  the kinematics  of
MaxBCG  clusters \citep{Koester2007}  and  inferred the  mean and  the
scatter of the mass-richness relation  (here richness is a measure for
the  number of  galaxies that  reside in  the cluster).   Becker \etal
combined  the  kinematics of  satellite  galaxies  in  finite bins  of
cluster richness  and measured  the second and  fourth moments  of the
host-weighted  velocity  distribution.  They  used  these two  moments
simultaneously  to   determine  the  mean  and  the   scatter  of  the
mass-richness  relation. This  method is  complementary to  that
presented here, and it will be interesting to compare both methods and
investigate  their relative  strengths  and weaknesses.  We intend  to
address this in a future study.

Finally we  emphasize that the scatter in  the conditional probability
function  $P(M|\lc)$ is expected  to increase  with increasing  $\lc$. 
This is due to the fact  that the slope of $\langle\lc\rangle(M)$,
which is  the mean  of $P(\lc|M)$,  becomes shallower  with 
increasing halo mass.  Hence, when stacking haloes  according to the
luminosity of the central galaxy,  one cannot ignore the  scatter in
$M$,  even when the scatter in  $P(\lc|M)$ is small.  This has
important  implications for any technique  that relies on  stacking,
such as  satellite kinematics and galaxy-galaxy lensing \citep[see
e.g.][]{Tasitsiomi2004,Cacciato2008}


\section*{Acknowledgments}

We are  grateful to Kris  Blindert, Anupreeta More, and  Peder Norberg
for valuable discussion.


\end{document}